\documentclass[conference]{IEEEtran}
\IEEEoverridecommandlockouts

\usepackage{amsmath,amssymb,amsfonts}
\usepackage{algorithmic}
\usepackage{graphicx}
\usepackage{subcaption}
\usepackage{textcomp}
\usepackage{xcolor}
 \usepackage[normalem]{ulem}
\usepackage{url}
\usepackage[style=numeric,sorting=none]{biblatex}
\def\BibTeX{{\rm B\kern-.05em{\sc i\kern-.025em b}\kern-.08em
    T\kern-.1667em\lower.7ex\hbox{E}\kern-.125emX}}
\begin{document}

\title{Boosting Performance for Software Defined Networks from Traffic Engineering Perspective\\
\thanks{}
}

\author{\IEEEauthorblockN{Mohammed I. Salman}
\IEEEauthorblockA{\textit{Department of Computer Science and Engineering} \\
\textit{Wright State University}\\
Dayton, Ohio \\
salman.6@wright.edu}
\and
\IEEEauthorblockN{Bin Wang}
\IEEEauthorblockA{\textit{Department of Computer Science and Engineering} \\
\textit{Wright State University}\\
Dayton, Ohio \\
bin.wang@wright.edu}
}

\maketitle

\begin{abstract}

Paths selection algorithms and rate adaptation objective functions are usually studied separately. In contrast, this paper evaluates some traffic engineering (TE) systems for software defined networking obtained by combining path selection techniques with average delay and load balancing, the two most popular TE objective functions. Based on TE simulation results, the best TE system suitable for software defined networks is a system where the paths are calculated using an oblivious routing model and its adaptation rate calculated using an average delay objective function. Thus, we propose the RACKE+AD system combining path sets computed using Räcke’s oblivious routing and a traffic splitting objective function using average delay. This model outperforms current state-of-the-art models, maximizes throughput, achieves better network resource utilization, and minimizes delay. The proposed system outperformed SMORE and SWAN by 4.2\% and 9.6\% respectively, achieving 27\% better utilization and delivering 34\% more traffic with 50\% less latency compared with both systems on a GÉANT network.

\end{abstract}


\section{Introduction}
Centralized traffic engineering (TE) has gained much attention following new software defined networking (SDN) developments. Large technology companies such as Microsoft \cite{SWAN} and Google \cite{B4} have shifted to this technology over the last few years.\par

Some previous studies have deviated from the standard SDN centralization feature to improve scalability and fast adaptation to changing traffic conditions, e.g. Contra \cite{contra}, HULA \cite{hula}, MP-HULA \cite{mp-hula}, and DASH \cite{dash} balance load traffic entirely in the data plane to reduce controller overhead. These solutions provide scalable systems with short response time, but degrade performance, with resulting distributed solutions far from optimal \cite{conga}.\par

Performance can also be affected by the traffic splitting objective function. Some TE systems balance load over some paths by minimizing maximum link utilization (MLU) \cite{SWAN, SMORE}. However, minimizing MLU does balance load and enhance performance for low traffic and degrades performance significantly during peak hours since it requires additional constraints to satisfy all the demands \cite{TED}. Other TE systems use meta-heuristic \cite{Tajiki2018} or heuristic \cite{8521584} solutions that can provide fast routing convergence, but the solutions are sub-optimal since they may be only local optima. Prior to SDN, several studies considered different objectives \cite{Gourdin06comparisonof, 10.1007/11753810_7}. To our knowledge, performance impacts from these objectives and path selection strategies have not been properly considered for SDN. Any TE system has two key ingredients: which set of paths is used for forwarding traffic, and how to split traffic over these selected paths. To the best of our knowledge, no previous study has focused on boosting performance by optimizing combinations of these key ingredients, in contrast, previous work has focused on either path selection algorithms or traffic-splitting objective functions, but not both.

Many studies suggest that a set of shortest paths should be used in TE systems to achieve reliable performance \cite{SWAN, 1039866, Ye2014}. Unfortunately, choosing shortest paths may exacerbate congestion for topologies with high link capacity heterogeneity. Oblivious routing\footnote{We use ``Oblivious routing" and ``Räcke's oblivious routing" interchangeably}  strategies offer network demand independent routing schemes, i.e., the routing scheme that is oblivious to the demands \cite{racke2003, cohen_applegate, racke2008, racke2020}. Although oblivious routing schemes can be devised with guaranteed congestion ratio, the resulting routing scheme is static and unable to adapt to changing traffic conditions. Several studies have shown that route allocations calculated using an oblivious routing model achieve comparable quality to adaptive solutions \cite{SMORE, coyote}. Selected paths from this approach are capacity-aware and diverse, which improves not only system performance, but also robustness. \par

The capacity aware concept not only applies to path selection only, but also to sending rates. For example, the Kleinrock delay objective function \cite{10.5555/1096922} minimizes congestion by increasing highly utilized link costs, thus, avoiding highly congested links. The widely used load balancing (LB) objective function \cite{SWAN,SMORE,Heorhiadi,X.Liu,MEDHI2018114} minimizes utilization (relative load) for all links, and can also be considered a capacity-aware objective function. The main goal for demand aware objectives is to mitigate proportional increases for all demands \cite{Gourdin06comparisonof} by minimizing MLU. However, all source destination (SD) pair demands do not increase at the same rate, and it is not trivial to predict future demands. Thus, sending rates should not only be capacity aware, but also demand aware.

Therefore, we constructed a new simulator, and motivated by SMORE \cite{SMORE} and AD objective functions \cite{X.Liu, MEDHI2018114, Bernard_Unsplittable} we propose RACKE+AD, a centralized, adaptive, semi-oblivious, demand aware, near optimal TE system with static routes allocated using Räcke’s oblivious routing model \cite{racke2003, racke2008, racke2020} and dynamic rate adaptation by approximating the average delay (AD) objective function. RACKE+AD outperformed SWAN \cite{SWAN} and SMORE \cite{SMORE} for throughput, congestion, and latency evaluated on GÉANT and ATT topologies.


\textbf{Contributions.} Critical contributions from the current paper are as follows:
\begin{enumerate}
    \item We present a routing scheme that outperforms current state-of-the-art techniques.
    \item We introduce RACKE+AD, a new efficient TE simulator that can test many routing schemes simultaneously. RACKE+AD is optimized for testing different route selection algorithm and objective function combinations and can be easily extended to test future TE systems. 
    \item We demonstrate that a TE system with static routes and adaptive traffic splitting offers many benefits, including performance, throughput, and resource utilization.
\end{enumerate}

\section{System Model}
All TE systems comprise two phases: identifying a set of paths to be used to forward traffic (path selection), and identifying splitting ratios to distribute traffic over these paths (rate adaptation). Generally, routes selected in the path selection phase are static, i.e., selected once and only recalculated when the network topology changes. Path selection is usually offline because updating end-to-end paths may take hundreds of seconds for wide area networks. In contrast, the rate adaptation phase must update path weights regularly due to frequent demand changes. However, the time required to update path weights is considerably less than the time required to update paths in the network. Among many techniques of paths selection algorithms and rate adaptation objective functions, the aim of this research is to find the best combination of these phases to enhance network performance.

\noindent \textbf{Path and Rate Adaptation Properties:}
Intuitively, independently chosen paths may not provide better performance than dependently chosen paths. However, SMORE showed that path selection has considerable effect on performance \cite{SMORE}. Selected paths should be low stretch to minimize latency and naturally load balanced to provide better performance. Low stretch motivated us to compare SMORE performance and latency against k-shortest paths (KSP) approaches. SMORE is naturally load balanced since route computation in Räcke’s oblivious routing model is not independent and incorporates some randomness, i.e., the obtained route set may not be the same if we were to run the model again. Thus, we expect different performance for each run. On the other hand, KSP selected paths are not capacity aware, whereas Räcke’s model selected paths are capacity-aware due to the natural load balancing. Performance can be further boosted if we use the same concept for splitting traffic over the selected paths, and we expect best performance may be achieved using phases, path selection, and rate adaptation.

\subsection{Rate Adaptation Models}

\subsubsection{Load Balance}

The load balance (LB) objective is also known as minimizing MLU, Wozencraft objective \cite{wozencraft}, or minimizing congestion, where LB minimizes the load on the most congested link. Thus, the LB problem can be expressed as \cite{MEDHI2018114}

\begin{small}
\begin{subequations}\label{LBequation:main}
\begin{align}
& \text{min}  && F(x) = r     &   & \tag{\ref{LBequation:main}} \\
& \text{s.t.} && \sum _{p \in P_{d} }^{}x_{dp}= h_{d},                                     & d & \in D \label{LBequation:d}  \\
&             && \sum_{d \in D~}^{} \sum_{p \in P_d}^{} \delta_{dpl} x_{dp} \leq c_l r, & l & \in L \label{LBequation:c}
\end{align}
\end{subequations}
\end{small}

where: $x_{dp}$ is the flow on path $p$ for demand $d$; $h_{d}$ is the volume for demand $d$; $ c_{l}$ is the capacity for link $l$; $ P_{d}$ is the number of candidate paths for demand $d$; $ \delta _{dpl}$ = 0, 1 is a link-path indicator, with $ \delta _{dpl}$ = 1 if path $p$ for demand $d$ uses link $l$, and 0 otherwise.

Two constraints are applied. The demand constraint \eqref{LBequation:d} ensures that all demands are satisfied over some paths. The capacity constraint \eqref{LBequation:c} ensures that load does not exceed the link capacity where $r \leq 1$, after solving \eqref{LBequation:main}. The linear program formulation above is the final form of the problem whereas the original problem is non-linear. The reader is referred to Chapter 4 of \cite{MEDHI2018114} for details on how the problem can be converted to the current form.

\subsubsection{Average Delay}

\begin{figure}[t]
  \centering
   \includegraphics[width=85mm]{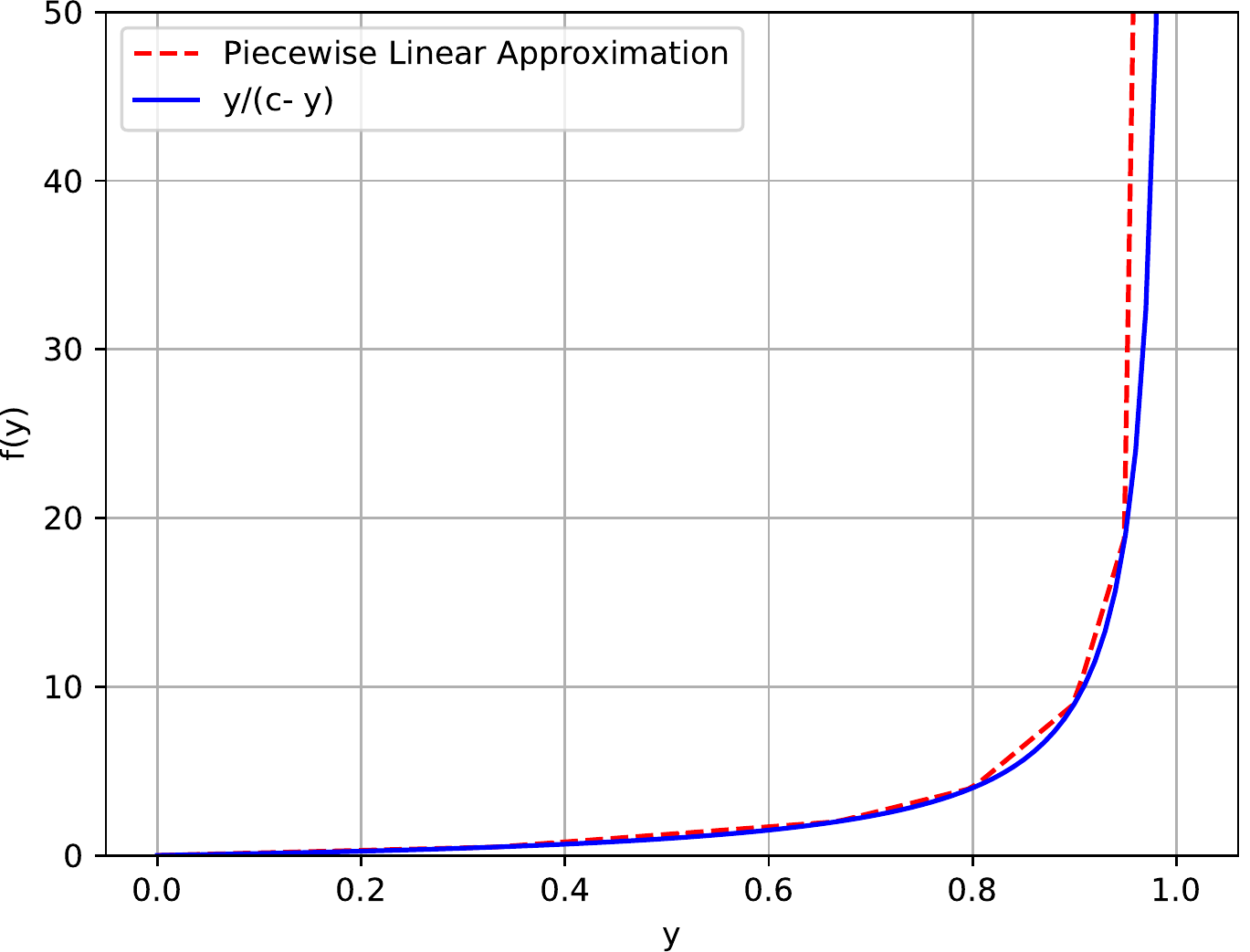}
  \caption{Piecewise linear approximation of the delay function.}
  \label{pla}
\end{figure}

For this objective function, delay for any network link can be modeled as $y/(c-y)$, as shown in (Figure~\ref{pla}, solid line). Similar to the LB objective, the original AD problem is non-linear and cannot be formulated directly as a linear program. Thus, the delay function is a piecewise linear approximation \eqref{ADequation} (Figure~\ref{pla}, dotted line)

\begin{small}
\begin{equation}\label{ADequation}
g(z) = \begin{cases}
(3/2)z &\text{for $1 \leq z < 1/3$}\\
(9/2)z-1 &\text{for $1/3 \leq z < 2/3$}\\
15z-8 &\text{for $2/3 \leq z < 4/5$}\\
50z-36 &\text{for $4/5 \leq z < 9/10$}\\
200z-171 &\text{for $9/10 \leq z < 19/20$}\\
4000z-3781 &\text{for $z \geq 19/20$}
\end{cases}
\end{equation}
\end{small}

The linear program for this AD problem is 

\begin{small}
\begin{subequations}\label{ADLP:main}
\begin{align}
& \text{min}  && F= \sum _{l=1}^{L}\frac{r_{l}}{c_{l}}     &   & \tag{\ref{ADLP:main}} \\
& \text{s.t.} && \sum _{p=1}^{P_{d}}x_{dp}=h_{d},                                    \quad d= 1, 2,...,D   \\
&             && \sum _{d=1}^{D} \sum _{p=1}^{P_{d}} \delta _{dpl} x_{dp}= y_{l},  \quad l= 1, 2,..., L  \\
&             && r_{l} \geq \frac{3}{2} y_{l}, 	   \quad l= 1, 2,..., L  \\
&			  && r_{l} \geq \frac{9}{2} y_{l}- c_{l}, \quad	 l= 1, 2,..., L  \\
&			  && r_{l} \geq 15 y_{l}- 8c_{l}, 		\quad l= 1, 2,..., L  \\
&			  && r_{l} \geq 50 y_{l}- 36c_{l},		\quad l= 1, 2,..., L  \\
& 			  && r_{l} \geq 200 y_{l}- 171c_{l}, 	\quad l= 1, 2,..., L  \\
&			  && r_{l} \geq 4000 y_{l}- 3781c_{l},	\quad l= 1, 2,..., L  \\
& 		   	  && x_{dp} \geq 0, 					\quad p=1,2,..., P_{k}, d=1, 2, ..., D \\
& 			  && y_{l} \geq 0, 						\quad l= 1, 2,..., L
\end{align}
\end{subequations}
\end{small}
which is considerably more accurate \cite{MEDHI2018114} than the Fortz et al. \cite{Bernard} approximation.

\subsection{Paths Selection Algorithms}
\subsubsection{Räcke's oblivious routing model}

Räcke’s oblivious routing model iteratively computes a distribution over randomized routing trees using an approximation algorithm. Link weights are adjusted for each iteration based on how much the link has been utilized in previous routing tree sets. A routing tree has leaves corresponding to nodes in the original topology. Thus, a path can be obtained between nodes $u$ and $v$ in the original graph by finding corresponding leaves for $u$ and $v$ in the routing tree. \par
However, paths for Räcke’s oblivious routing model are computed without considering demands, thus, they do not overfit to a specific scenario \cite{SMORE}. Similar to SMORE, we also adopt the simple mechanism used to impose the number of paths for each SD node pair. We use 4 paths for each SD pair of nodes that have the highest weights.

\subsubsection{\textit{K-shortest} paths}
The proposed KSP algorithm is based on Yen’s algorithm, the most commonly used algorithm for TE. KSP is a generalization of the shortest path routing problem. The algorithm returns loopless $k$ shortest paths ordered from shortest to longest. We use four paths for each SD pair, i.e., $k = 4$.

\section{Simulator Framework}
We built a simulator to model and test different TE scenarios, with particular attention to efficiency, simplicity, and extendibility. Although many network simulators have been proposed previously \cite{NS-3, OPNET, YATES, REPETITA}, they are generally not optimized for modeling TE approaches and/or do not provide ease of use or extendibility. The proposed simulator was built in Python and can test many TE models in parallel while recording statistics in the background. We use Gurobi optimization \cite{gurobi} to solve the linear programming problems, by integrating it with Python. The framework, data and Räcke’s oblivious routing model implementation are all available online\footnote{\url{https://github.com/MohammedSalman/TE-SIMULATOR}}. \par
Simulator inputs, (e.g. topology, demands, path selection algorithms, objective functions, etc.) are all specified in a Python script or configuration file. The simulation produces visualized throughput graphs for each TE system. The graphs are updated periodically as throughput data becomes available. Three time-series metrics for each TE system are recorded in the background during simulation: overall throughput, congestion per link, and latency per path. Topology and traffic matrices are provided as input files, where the user provides the location to these files in the configuration file. If the locations are unavailable, random topology and traffic matrices will be generated according to provided parameters, including number of nodes $N$, number of links $L$, and traffic distribution matrix.

\section{Simulation Setup}

\subsection{Evaluating Routing Scheme Quality}

We evaluate TE systems based on congestion, throughput, and delay. Congestion reflects how a TE system utilizes network resources, and we mostly care about congestion when traffic demand exceeds link capacity. Thus, avoiding congestion can be considered as preserving as much residual capacity as possible, which is important for unexpected traffic surges that could cause bottlenecks. Congestion has negative impact on delay due to queuing. We measure path delay by summing queuing delay for each link along that path, $l/(c-l)$, where $l$ is the absolute link load and $c$ is the link capacity. Throughput is the proportion of total demand that is successfully delivered to the destinations.

\subsection{Simulation Settings}
\noindent \textbf{Path selection algorithms.} We use three approaches for path selection  (i) paths selected using Räcke’s oblivious routing model, (ii) paths selected using KSP algorithm, and  (iii) select all available simple paths. We refer to these RACKE, KSP, and OPTIMAL, respectively. \par 
\noindent \textbf{Rate adaptation objective functions.} We use two objective functions for rate adaptation: AD and LB. We refer to a routing scheme with paths selected using KSP and rate adaptation using LB objective function as KSP+LB. Similarly, models where the routing scheme  selects all available paths and rate adaptation uses AD is referred to as OPTIMAL (AD), etc. The RACKE+LB routing scheme parallels that used in SMORE \cite{SMORE}, and KSP+LB is an approximation to the SWAN scheme \cite{SWAN}. Table \ref{TE_SYSTEMS} shows the TE systems used in our experiment.\par

\begin{table}[h]
\caption{Implemented TE algorithms }
\label{TE_SYSTEMS}
\begin{center}
\begin{minipage}{9cm}
\begin{tabular}{|l||l|}
\hline
TE System & Description\\
\hline
KSP+LB & \textit{k-Shortest} Paths (KSP) for paths, LB for weights\\
KSP+AD & \textit{k-Shortest} Paths (KSP) for paths, AD for weights\\
RACKE+LB & Räcke’s oblivious routing for paths, LB for weights\\
RACKE+AD & Räcke’s oblivious routing for paths, AD for weights\\
OPTIMAL(LB)\footnote{The best load balance is achieved with this system.} & All paths, LB for weights \\
OPTIMAL(AD)\footnote{The best average delay is achieved with this system.} & All paths, AD for weights \\
\hline
\end{tabular}
\end{minipage}
\end{center}
\end{table}

\noindent \textbf{Path budget.} Similar to SMORE and SWAN, and to ensure a fair comparison, we use 4 paths to evaluate any routing scheme. If the Räcke’s oblivious routing model produces a routing scheme with SD pairs that has more than 4 paths, we use the 4 highest weight paths, similar to SMORE. \par

\noindent \textbf{Traffic matrix generation.} We use the gravity model to generate the traffic matrix (TM) \cite{SMORE,cohen_applegate}. The gravity model approximates real-world TMs for a production network \cite{10.1145/1096536.1096551}. TMs are deduced based on incoming/outgoing flow for each forwarding device. Since that information is not available, we use a capacity based heuristic rather than incoming/outgoing flow information \cite{cohen_applegate}. \par

\begin{figure}[htbp]
\centerline{\includegraphics[scale=0.25]{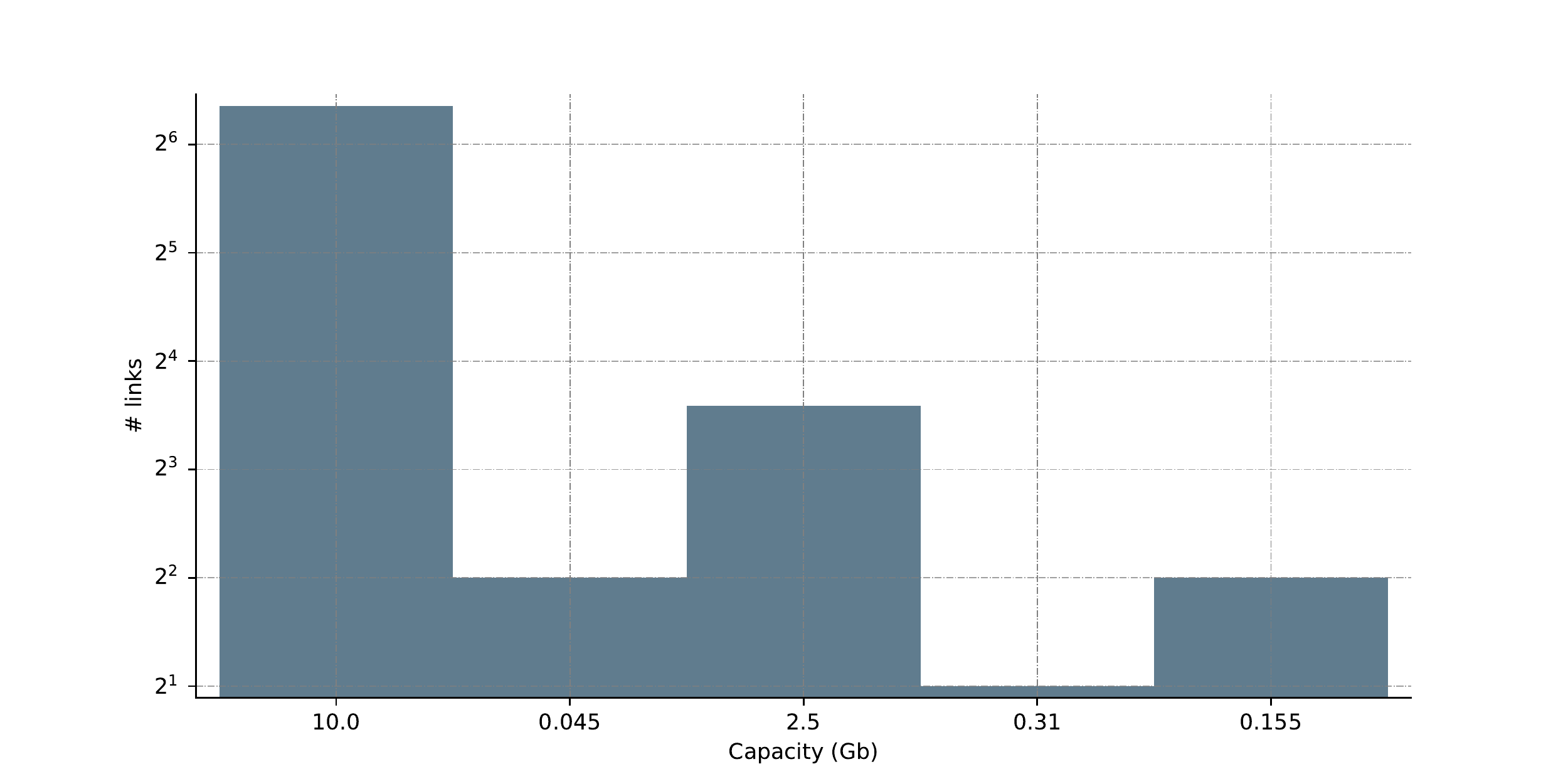}}
\caption{Capacity distribution for GÉANT network (log scaled).}
\label{_capacity_distribution}
\end{figure}

\noindent \textbf{Topologies.} We evaluate many TE systems for ATT and GÉANT\footnote{dataset available at: http://www.topology-zoo.org/dataset.html} production topologies. The GÉANT network (European academic network) contains 38 nodes and 104 directed links with heterogeneous capacities. Fig.~\ref{_capacity_distribution} shows the link capacity distribution for this network. Different TE systems may behave differently depending on link capacity distributions. Shortest-path TE systems may introduce a bottleneck in heterogeneous link capacities as many SD pairs compete for the same resources.

\section{Results}
We evaluated multiple routing schemes using criteria focused on: 
\begin{itemize}
    \item how each TE system performs regarding throughput and congestion, and
    \item SMORE and  KSP TE system impacts on latency.
\end{itemize}

\subsection{Throughput}
\begin{figure}[!ht]
 \centering
 \begin{subfigure}[b]{0.50\textwidth}
 \centering
  \includegraphics[width=1\textwidth]{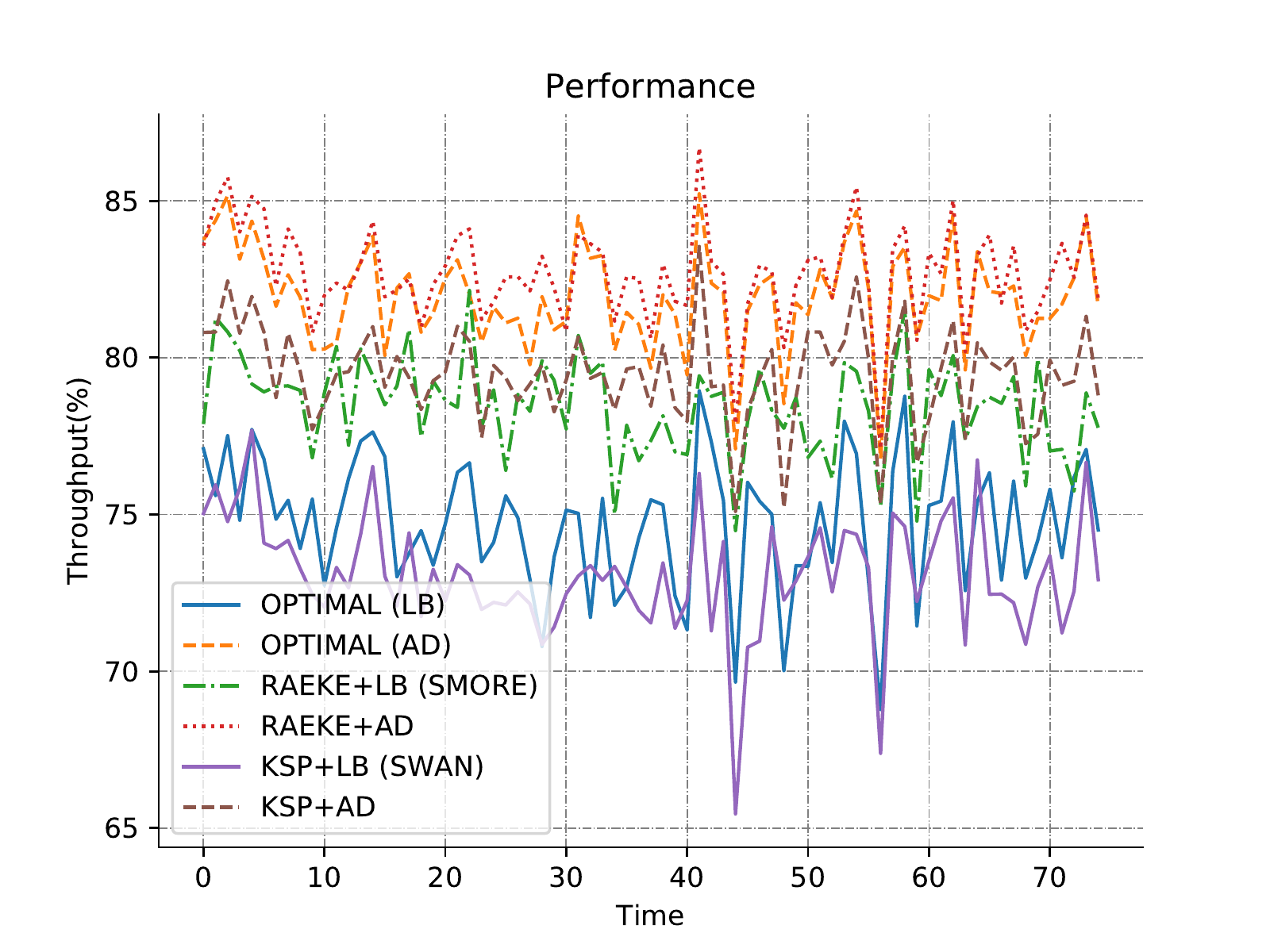}
  \caption{Throughput}\label{geant_throughput:throughput}
 \end{subfigure}\hfill
 \begin{subfigure}[b]{0.50\textwidth}
 \centering
  \includegraphics[width=1\textwidth]{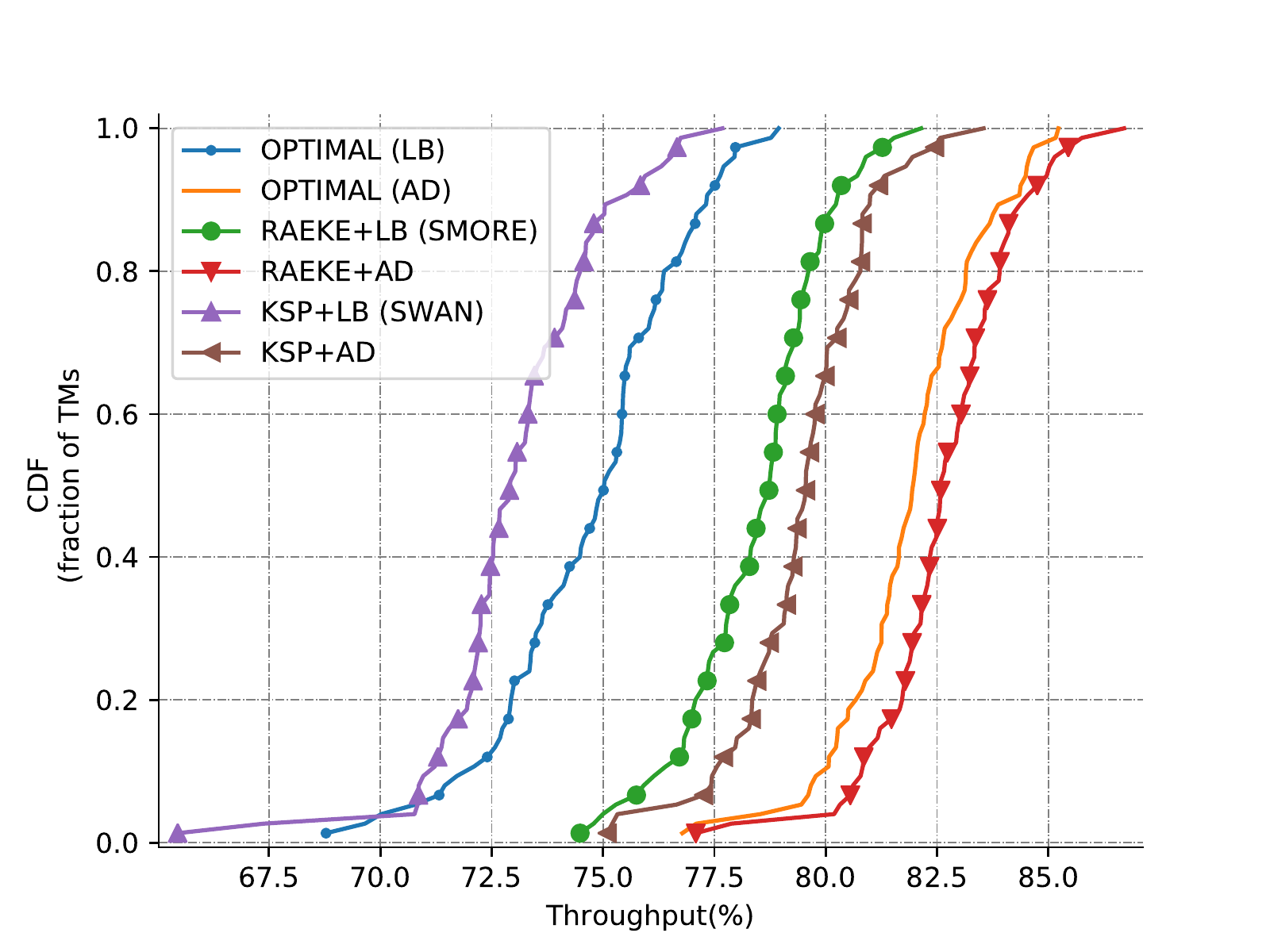}
  \caption{Throughput distribution}\label{geant_throughput:throughput_distribution}
 \end{subfigure}\hfill
 \caption{Throughput for GÉANT topology}\label{geant_throughput}
\end{figure}

Performance for many TE systems were evaluated on GÉANT and ATT networks with path budget = 4 for a fair comparison with SMORE. Figures ~\ref{geant_throughput:throughput} and ~\ref{geant_throughput:throughput_distribution} show throughput and corresponding throughput distribution for GÉANT network, respectively. Rate adaptation using AD objective function significantly increases throughput, achieving 4.2\% and 9.6\% improvement over SMORE and KSP+LB, respectively, which confirms path selection effectiveness using Räcke’s oblivious routing algorithm.

\begin{figure}[!ht]
\centering
        \begin{subfigure}[b]{0.50\textwidth}
            \centering
            \includegraphics[width=1\textwidth]{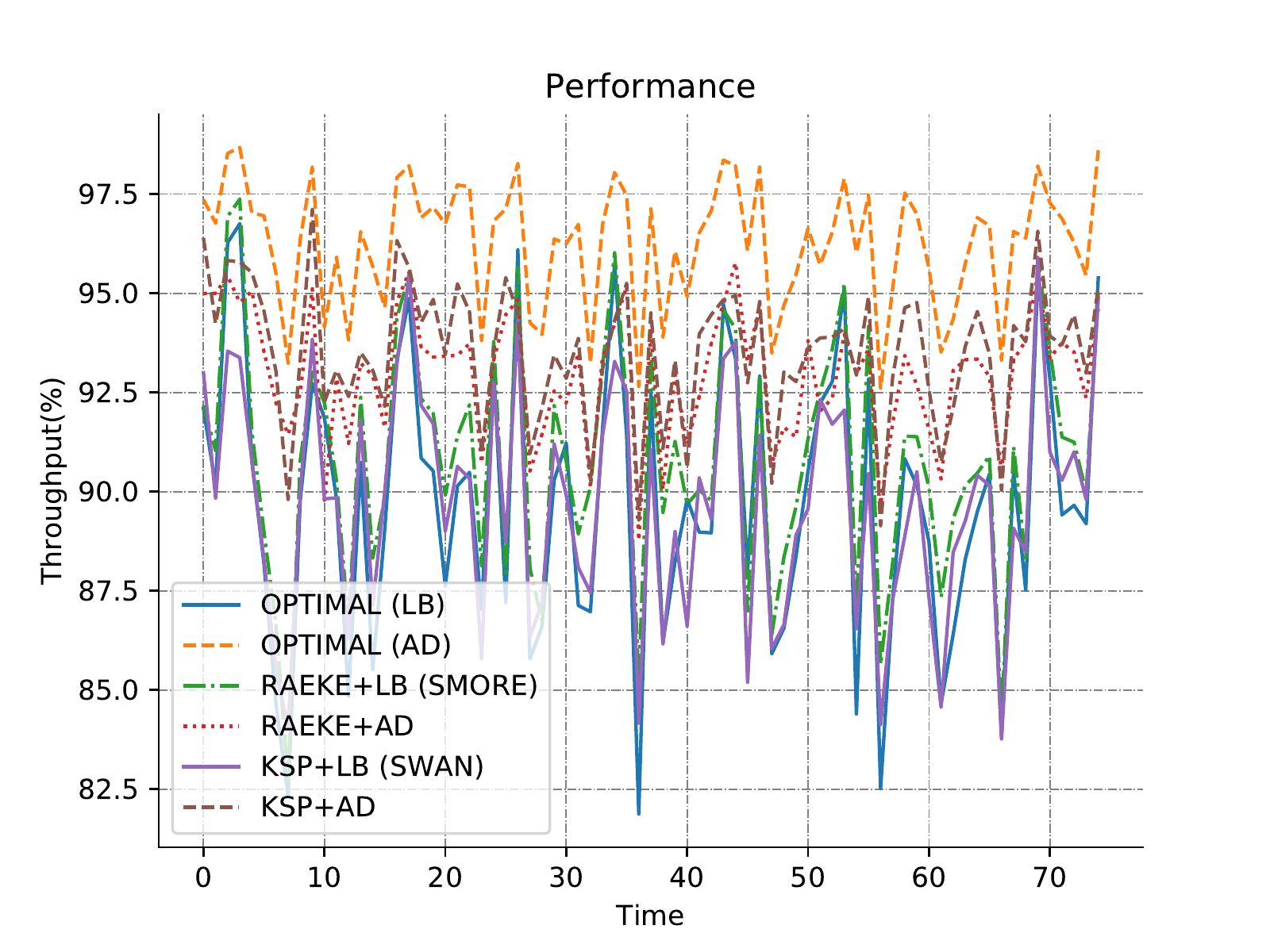}
            \caption{Throughput}\label{att_throughput:throughput}
\end{subfigure}\hfill
        \begin{subfigure}[b]{0.50\textwidth}
        \centering
        \includegraphics[width=1\textwidth]{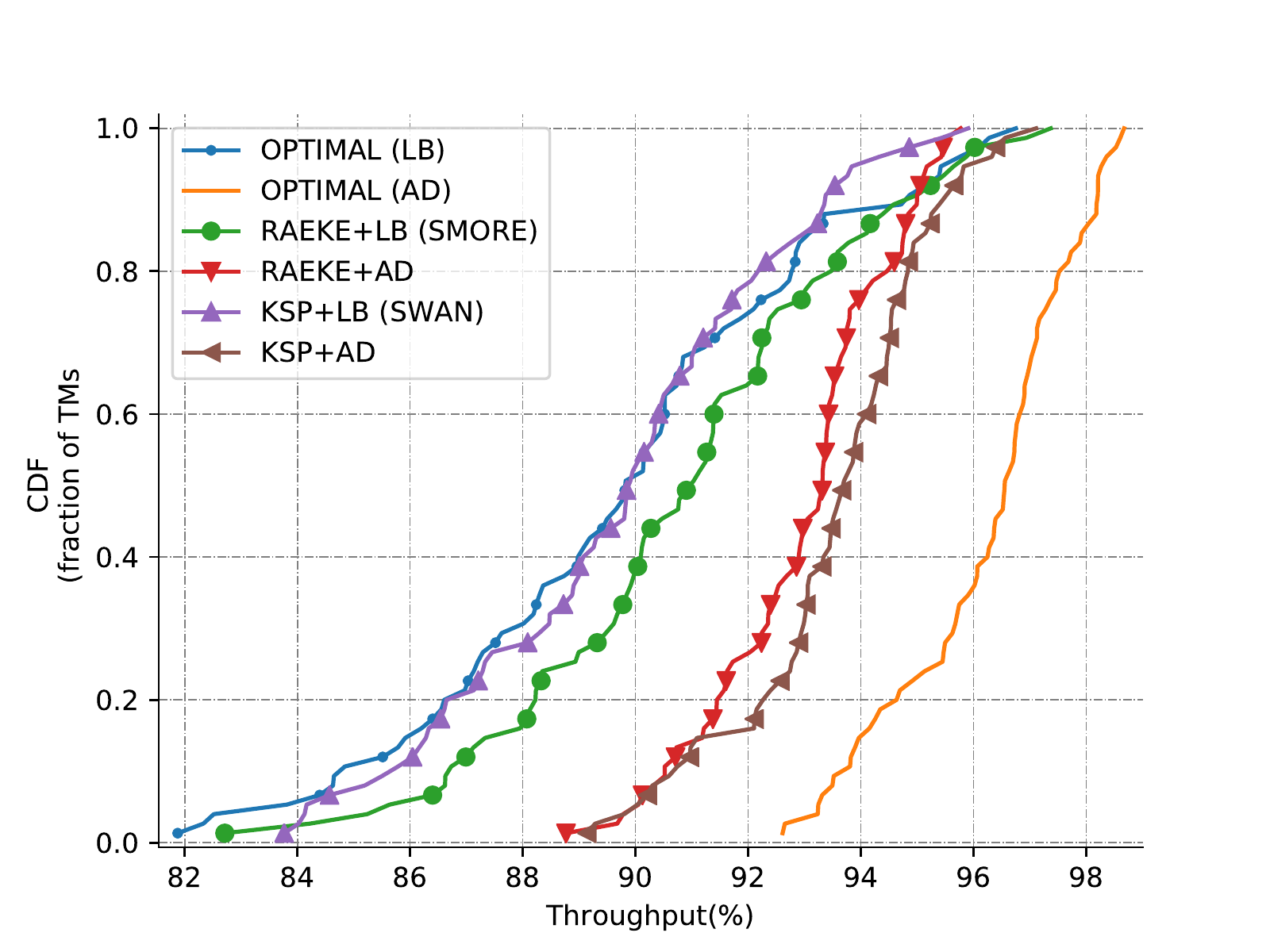}
  \caption{Throughput distribution}\label{att_throughput:throughput_distribution}
\end{subfigure}\hfill
 \caption{Throughput on ATT topology}\label{att_throughput}
\end{figure}

Similar to GÉANT topology, a higher throughput was achieved for ATT topology using the AD adaptation rate objective function. KSP had slightly better throughput than Räcke’s oblivious routing path selection algorithm (Figs.~\ref{att_throughput:throughput} and ~\ref{att_throughput:throughput_distribution}). \par

\begin{figure}[b]
\centerline{\includegraphics[scale=0.55]{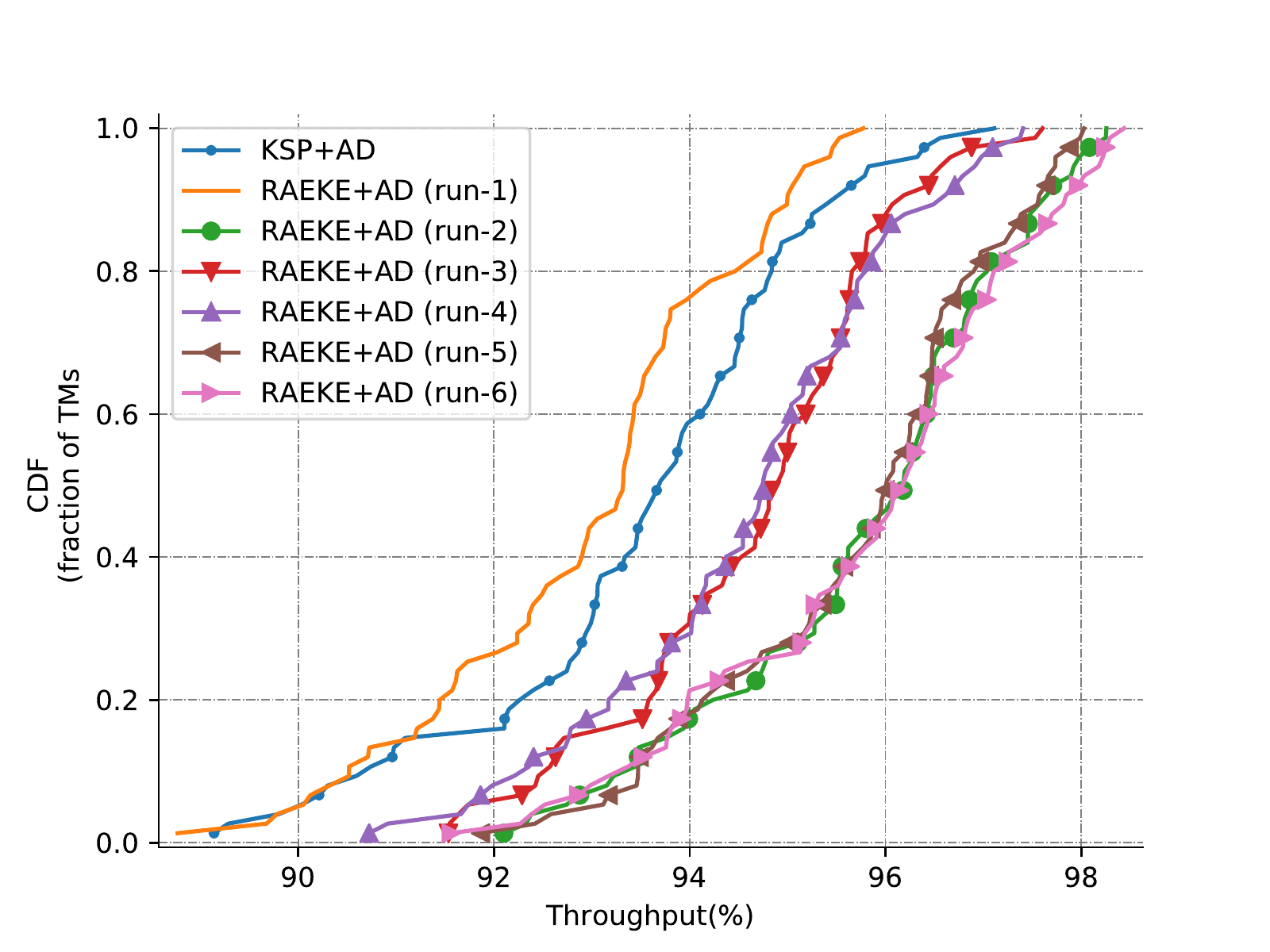}}
\caption{Throughput distribution for ATT topology for 1 KSP and 6 Räcke  schemes}
\label{throughput_distribution_many_runs}
\end{figure}

Räcke’s oblivious routing model with LB adaptation rate performed 1.14\% better than KSP on average. This may confirm that AD favors shortest paths when all links have the same capacity. However, there is no guarantee that SMORE will always outperform (or underperform) KSP under the same conditions due to oblivious routing scheme randomness. Figure ~\ref{throughput_distribution_many_runs} shows throughput distributions for KSP+AD with a different Räcke’s oblivious routing TE systems obtained by repeatedly calculating the oblivious routing scheme. Output from KSP+AD remained constant since KSP+AD is deterministic. Räcke’s oblivious routing scheme outperformed KSP for 5 runs and underperformed for 1 run. Thus, there is a worst case scenario where KSP may perform better than SMORE. The best run had 2.29\% higher throughput than KSP+AD. Therefore, a network operator may choose to run Räcke’s scheme several times and choose the best outcome.

\subsection{Congestion}
\begin{figure}
 \centering
 \begin{subfigure}[b]{0.50\textwidth}
 \centering
  \includegraphics[width=1\textwidth]{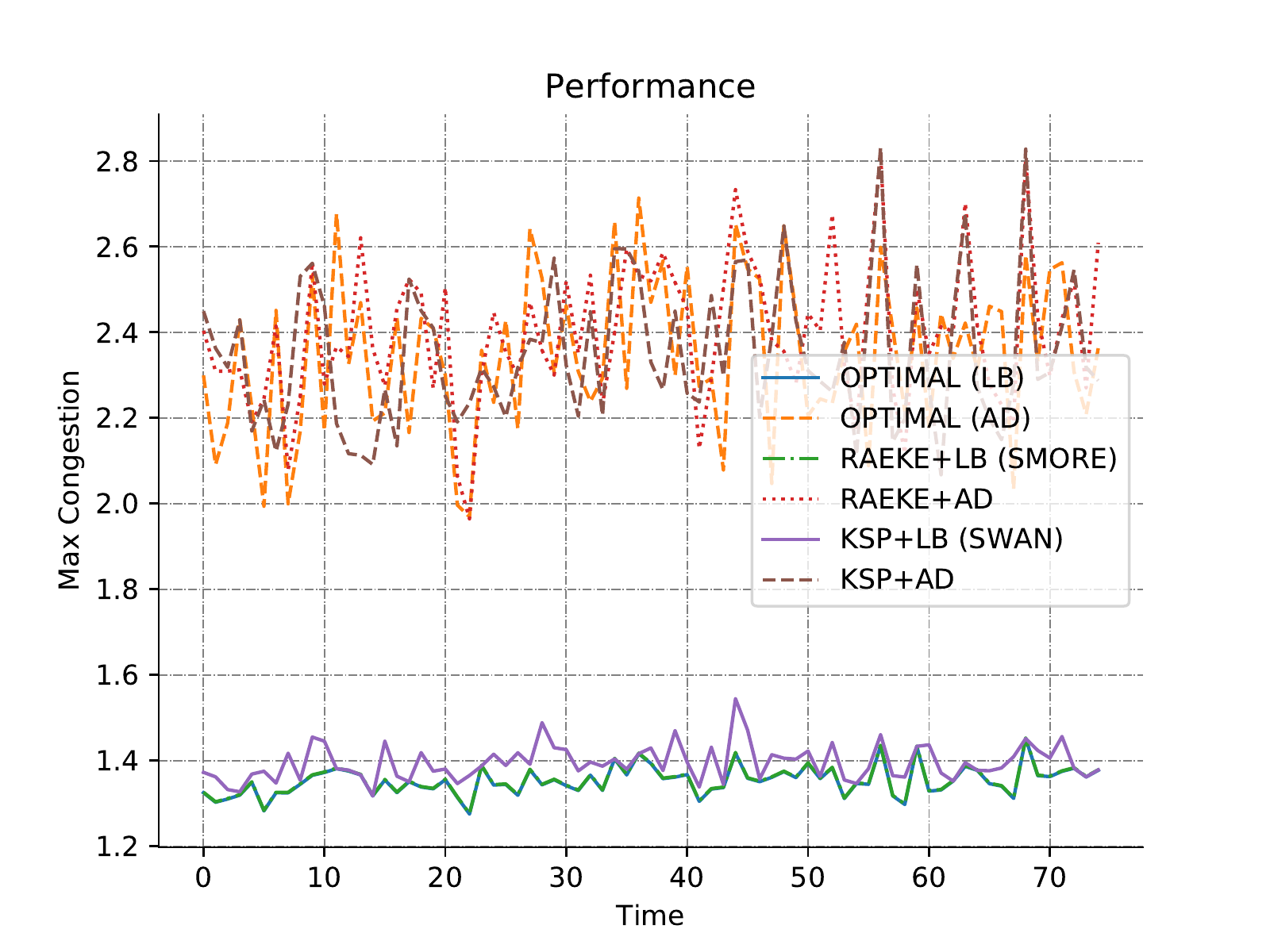}
  \caption{Max link congestion, GÉANT topology}\label{geant_congestion:max_congestion}
 \end{subfigure}\hfill
 \begin{subfigure}[b]{0.50\textwidth}
 \centering
  \includegraphics[width=1\textwidth]{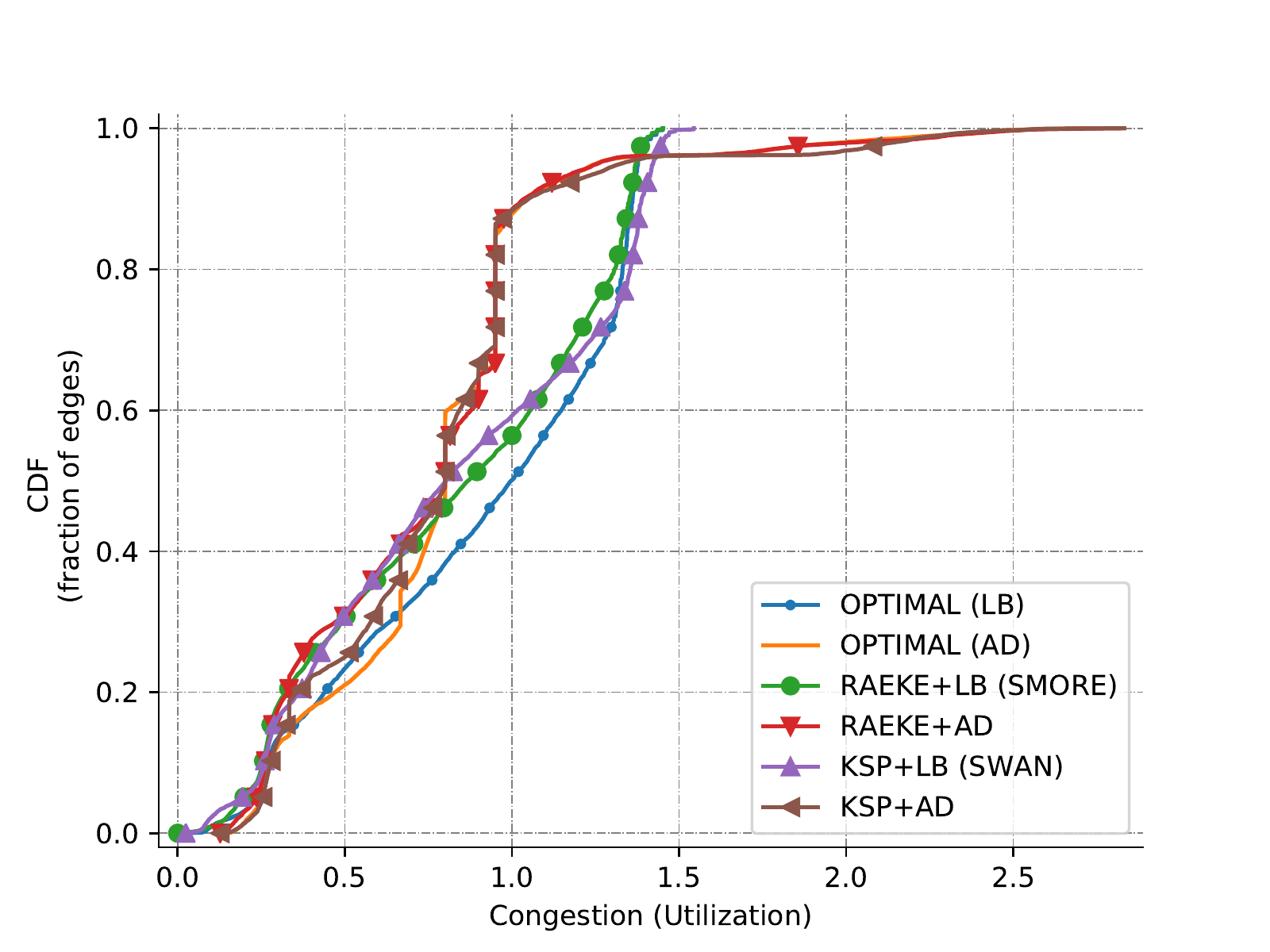}
  \caption{CDF of link congestions, GÉANT topology}\label{geant_congestion:congestion_distribution}
 \end{subfigure}\hfill
 \caption{Max link congestion and links' congestion distribution on GÉANT topology}\label{geant_congestion}
\end{figure}

Figures ~\ref{geant_congestion:max_congestion} and ~\ref{geant_congestion:congestion_distribution} show network congestion for GÉANT topology using AD and LB. The AD objective function scheduled link loading differently from LB. Figure ~\ref{geant_congestion:max_congestion} shows the maximum congested link over time. All TE systems scheduled link loads that exceeded specific link capacities since we deliberately fed the system with high volume demands to investigate TE system performance well under stressed conditions. AD (Fig.~\ref{geant_congestion:max_congestion}) seems to have higher MLU whereas Fig. ~\ref{geant_congestion:congestion_distribution}) shows that the AD objective utilizes link loads much better than LB. TE systems with LB caused a bottleneck for more than 40\% of links whereas TE systems with AD objective caused a bottleneck for 13\% of links. This low congestion ratio for AD is the main reason for the higher throughput (Fig.~\ref{geant_throughput}). \par
The LB objective always distributes traffic perfectly across the available routes, in the sense that all paths are used and all nodes send and receive traffic with quite similar link utilization (relative load) for all links. Thus, all links might be over-utilized under high demands when the system is not feasible. On the other hand, AD deals more with delay and throughput, but generates worse MLU than from LB. However, MLU is not a true network metric as it only considers congestion for a single link rather than the whole network. Thus, congestion distribution seems like a more reasonable metric, and we only measured MLU to make that point since it is heavily used in the literature. \par
\begin{figure}[!ht]
 \centering
\begin{subfigure}[b]{0.50\textwidth}
\centering
  \includegraphics[width=1\textwidth]{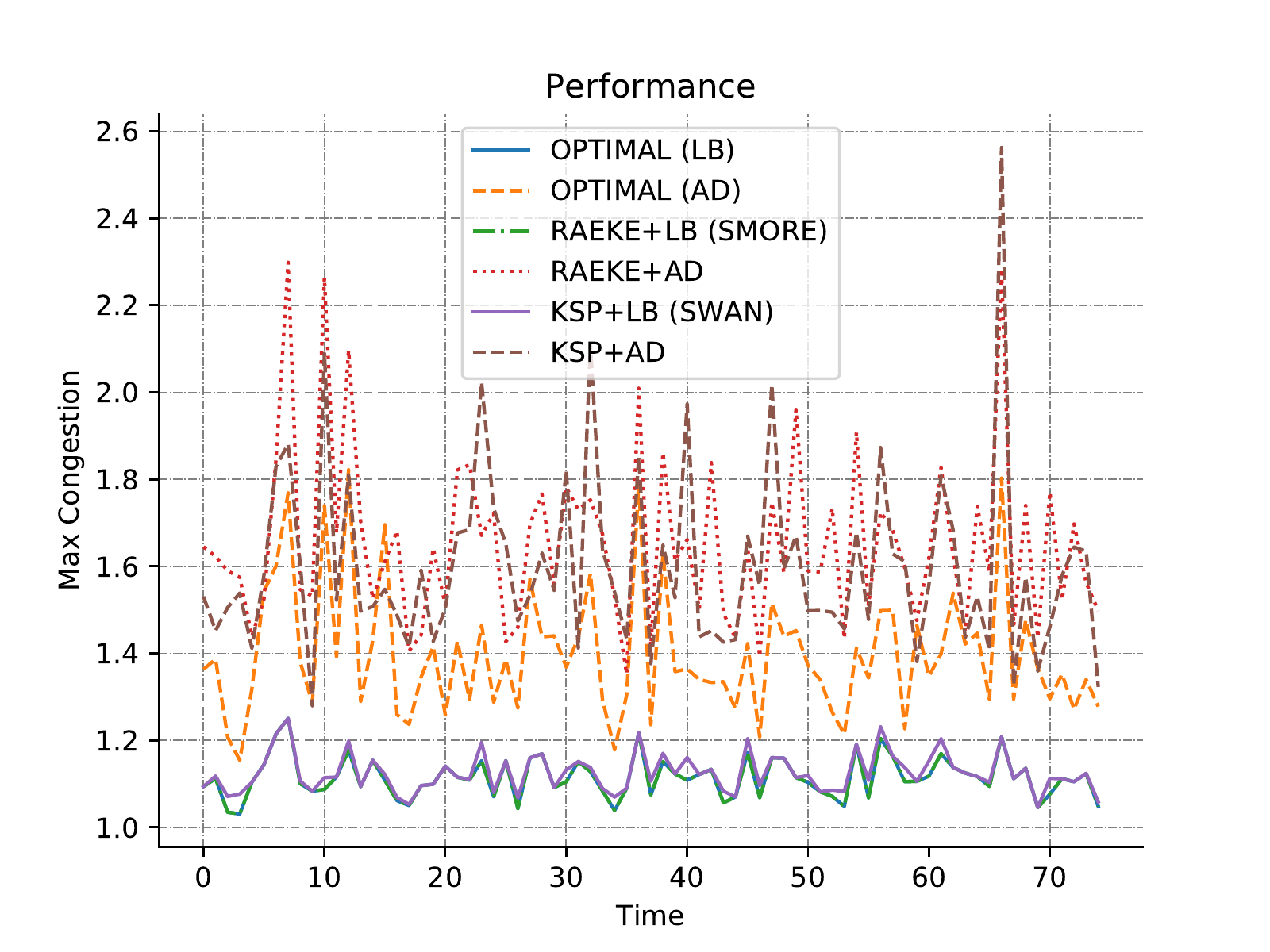}
  \caption{Max link congestion, ATT topology}\label{att_congestion:max_congestion}
 \end{subfigure}\hfill
 \begin{subfigure}[b]{0.50\textwidth}
 \centering
  \includegraphics[width=1\textwidth]{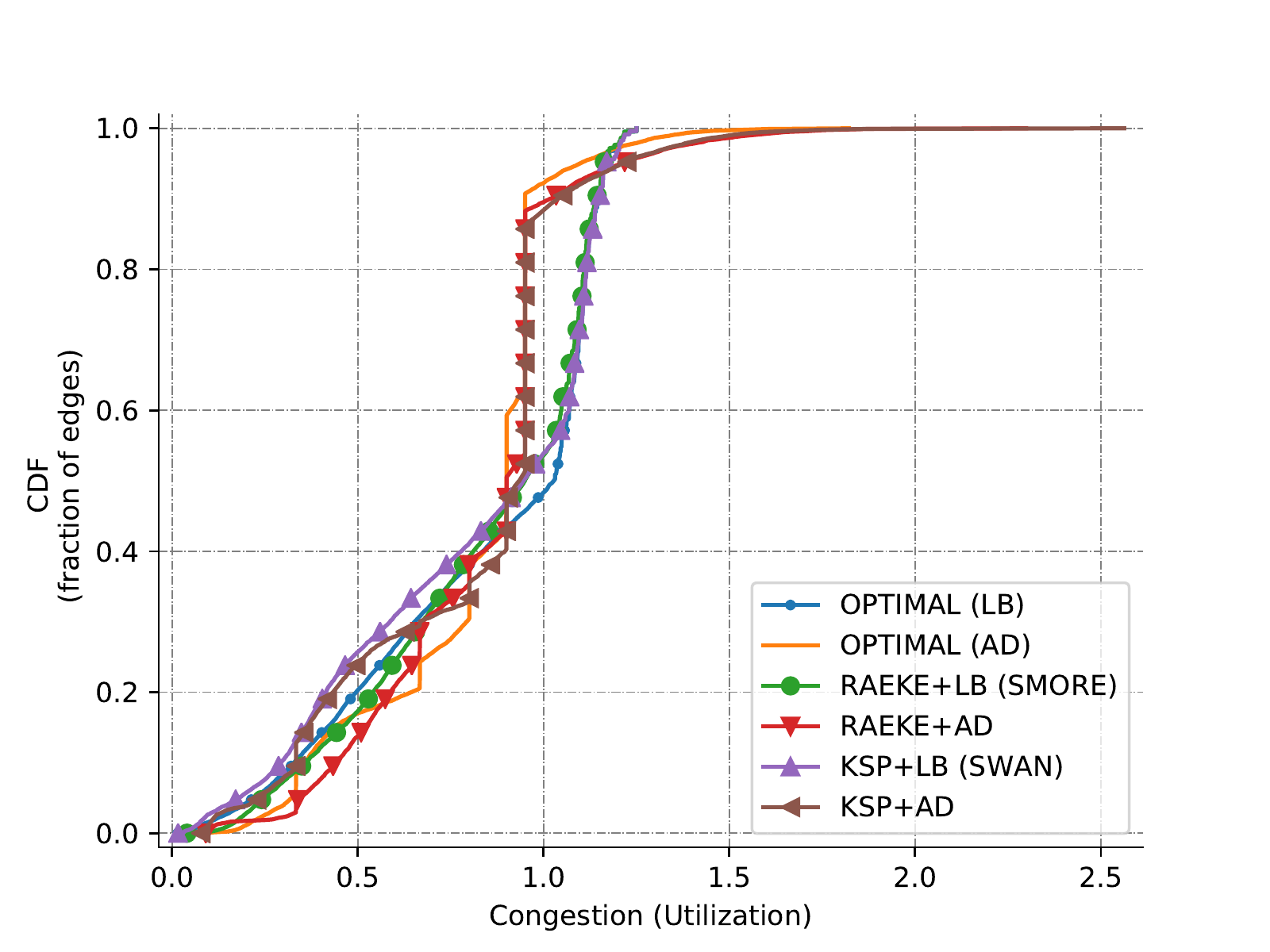}
  \caption{CDF of link congestions, ATT topology}\label{att_congestion:congestion_distribution}
 \end{subfigure}\hfill
 \caption{Max link congestion and links' congestion distribution on ATT topology}\label{att_congestion}
\end{figure}

Thus, two factors contributed to better throughput and less congestion: routes selected using Räcke’s oblivious routing algorithm, and using the AD objective. Similar results were obtained for ATT topology (Figs. ~\ref{att_congestion:max_congestion} and  ~\ref{att_congestion:congestion_distribution}). \par

\par

\subsection{Latency}
\begin{figure}[!ht]
 \centering
 \begin{subfigure}[b]{0.50\textwidth}
  \centering
  \includegraphics[width=1\textwidth]{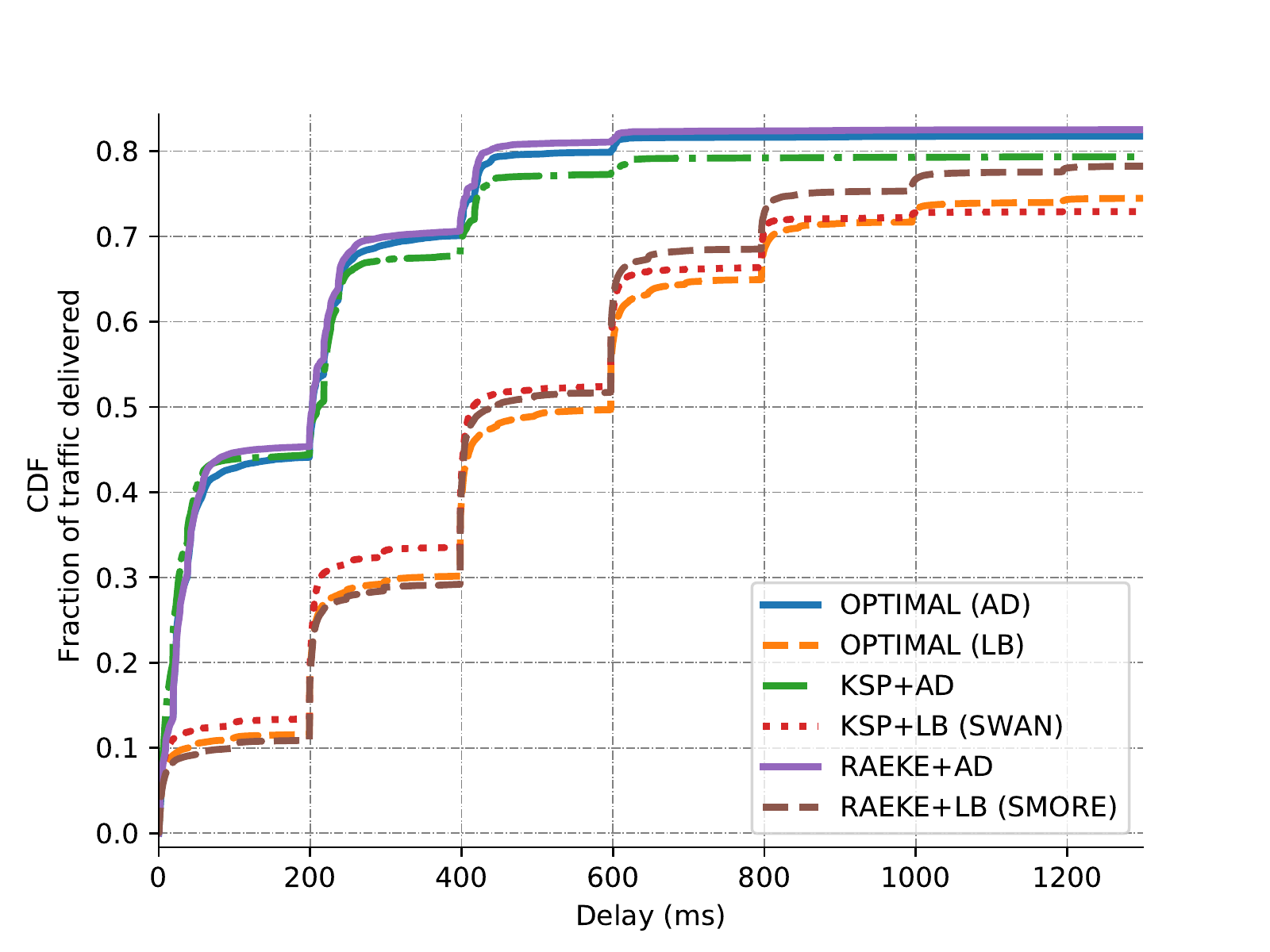}
  \caption{GÉANT topology.}\label{latency_distribution:geant}
 \end{subfigure}\hfill
 \begin{subfigure}[b]{0.50\textwidth}
 \centering
  \includegraphics[width=1\textwidth]{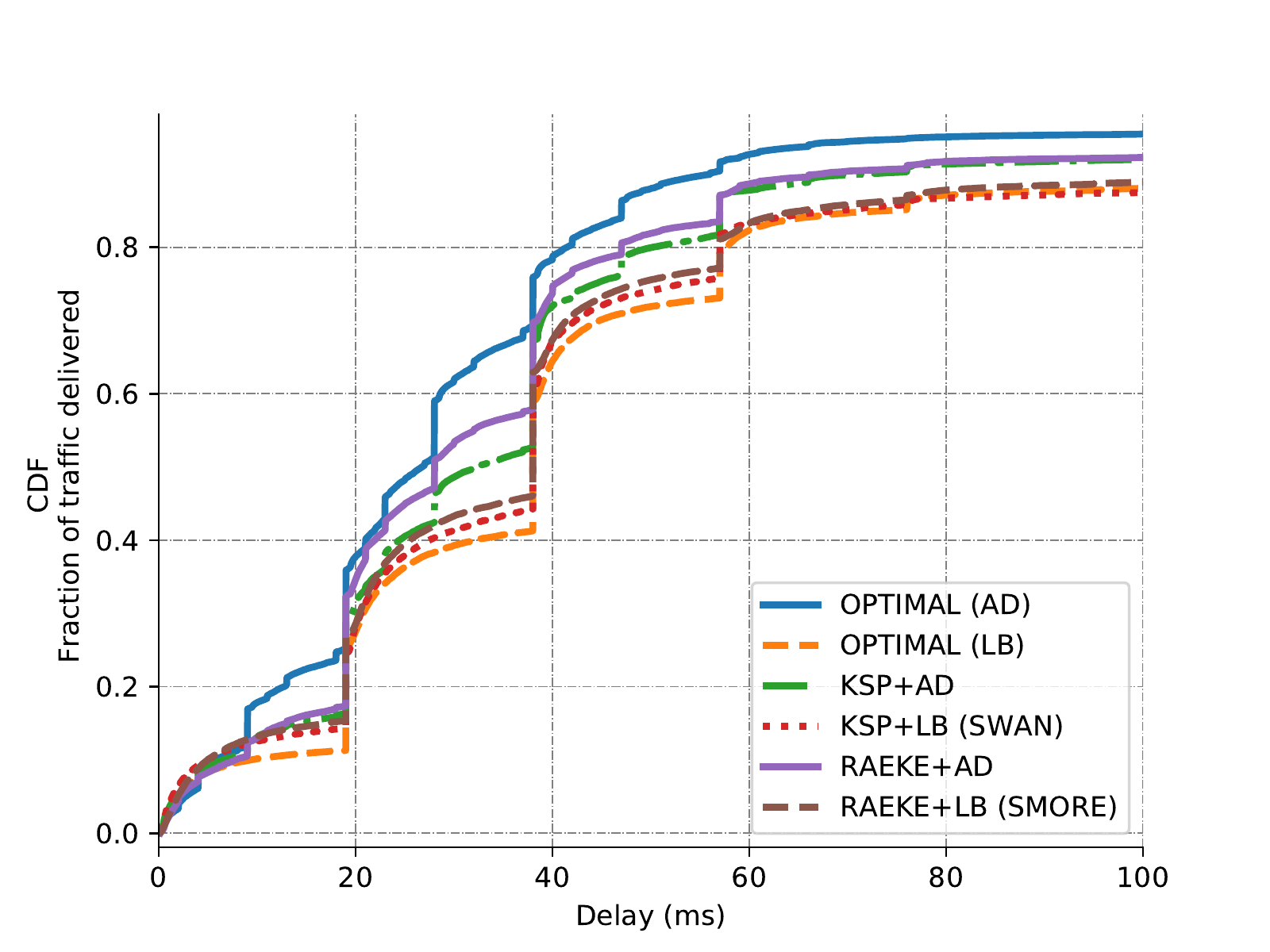}
  \caption{ATT topology.}\label{latency_distribution:att}
 \end{subfigure}\hfill
 \caption{Latency distribution}\label{latency_distribution}
\end{figure}

Figure ~\ref{latency_distribution} shows link delay distribution with respect to traffic delivered within that delay for GÉANT and ATT topologies. Latency for each path was computed by summing the link delays to obtain the path delay. Including AD selection outperforms LB, achieving significantly lower latency. 
Figure ~\ref{latency_distribution:geant} shows that LB and AD TE systems different considerably for GÉANT topology. TE systems with AD objective initially deliver approximately 34\% traffic more than those with LB objective, which also has latency 50\% lower latency than TE systems with AD. RACKE+AD routing delivered slightly more traffic than OPTIMAL(AD) since OPTIMAL(AD) goal is to reduce total delay rather than throughput. 
Figure ~\ref{latency_distribution:att} shows that routing schemes with AD also delivered more traffic than those with LB for ATT topology. However, the gap between the two groups is somewhat smaller than for GÉANT topologies (Fig. 8(a)) because ATT network links are heterogeneous, hence smaller performance differences between individual links.

\section{Related Work}\label{Related-work}

The classic approach for TE problems is to solve them as a linear program (LP) \cite{MEDHI2018114,wozencraft}, referred to as a multi-commodity flow problem, where the objective function usually minimizes MLU. The approximation of AD objective function is not as widely as used. However, this classical approach does not consider decoupling TE system phases because all available paths are provided as inputs. Choosing all available paths has two limitations: more paths means more decision variables in the LP, and forwarding devices, such routers and switches, have limited TCAM memory, hence fewer number paths is always preferable to keep the routing table as small as possible.\par

The conventional approach adjusts link weights to find a good routing scheme that can increase throughput or minimize congestion in the network \cite{Bernard, sculpte}. However, OSPF can never reach optimal because it uses the equal cost multi-path approach that splits traffic evenly among available shortest paths without rate adaptation. Furthermore, optimizing link weights is an NP-hard problem. \par

Potentially centralized TE approaches recently became viable due to software-defined networking (SDN) developments, that clearly decouple the two TE phases. SWAN \cite{SWAN} distributes traffic over a set of k-shortest paths using an LP that reserves a small amount of scratch capacity on links to apply updates in a congestion-free manner. SOL \cite{Heorhiadi} uses a greedy approach to randomly select paths with the promise that this random selection will help load balancing traffic across the network. This latter approach is somewhat similar to valiant load balancing \cite{VLB} but can lead to unnecessarily long paths and consequently increased latency. \par

Oblivious routing \cite{racke2003, cohen_applegate,racke2008} has also been proposed to find a routing scheme that performs well under all possible demands. The Räcke oblivious routing model \cite{racke2003} guarantees a congestion rate that is never worse than \textit{O}(log \textit{n}) of optimal, where $n$ is the number of nodes in the graph. However, despite the guaranteed congestion ratio, this approach cannot outperform systems like SWAN since it considers all possible traffic demands. On the other hand, the oblivious routing approach has inspired several studies (including the current study) to investigate a careful path selection approach.
SMORE \cite{SMORE} was inspired by Räcke’s oblivious routing model to carefully select paths that increase TE system performance and robustness. Paths selected this way have low stretch, which is important to decrease latency, and are capacity aware, which is important for load balancing.
The proposed approach in this paper suggests that careful route selection is not sufficient performance enhancement to reach the expected maximum performance. However, a different objective function from the commonly employed LB could further enhance performance. Hence we were inspired to compare LB and AD objective function performance, and subsequently propose the RACKE+AD TE system using oblivious routing for path selection with AD to achieve better link delay and network performance.


\section{Discussion}
This section discusses the reason behind the high gap in performance and delay between LB and AD objective functions and one potential limitation for this work. The LB objective function tends to make the relative load the same for all links when all SD pairs are sending and receiving traffic. This can enhance performance to some extent but causes bottlenecks between some SD pairs under stressed conditions and unpredicted demands, with consequential congestion loss. On the other hand, the AD objective function increases the cost for highly utilized links to avoid utilizing them if other less heavily utilized links are available. Thus, AD is more demand aware than LB and hence offers better contribution to performance. However, solving LP for LB is much faster than for AD, particularly for larger networks due to the increased number of constraints and decision variables.

\section{Conclusion}
Although a few TE systems have been optimized previously using different path selection algorithms, few studies have investigates performance enhancement by testing many objective functions for splitting traffic. These phases have only been studied in isolation previously, with no prior studies testing all possible combinations to find a routing scheme with the best available performance.\par
This paper proposed RACKE+AD TE system and validated its performance advantages by testing many possible combinations. RACKE+AD selects routes using Räcke’s oblivious routing model and the average delay objective function. Although the intuitive AD goal is to minimize network delay, it also provides surprisingly better throughput than minimizing MLU (commonly known as load balancing).\par
Simulations confirmed the proposed RACKE+AD system outperformed state-of-the-art routing TE systems in terms of throughput, congestion, and delay. We discussed a caveat when running Räcke’s oblivious routing model, where k-shortest paths may give better performance due to randomness in oblivious routing, and also discussed the importance of excluding the maximum congestion metric when evaluating TE systems, particularly system that split traffic not based on the LB objective function.

\section*{Acknowledgment}

We would like to thank the anonymous reviewers for their helpful comments and suggestions. We also would like to thank Praveen Kumar from Cornell University for addressing all the questions we had regarding the SMORE traffic engineering system.
\printbibliography

\end{document}